# The de Haas–van Alphen Effect Study of the Fermi Surface of ZrB$_{12}$

V.A.Gasparov [1], I. Sheikin [2], F. Levy [2], J. Teyssier [3], G. Santi [3]

1. *Institute of Solid State Physics RAS, Chernogolovka, 142432, Russian Federation*
2. *Grenoble High Magnetic Field Laboratory, CNRS, BP166, 38042 Grenoble Cedex 9, France*
3. *Département de Physique de la Matière Condensée, Université de Genève, 1211 Genève 4, Switzerland*

The de Haas–van Alphen (dHvA) effect in the cluster superconductor ZrB$_{12}$ was studied by magnetic torque measurements in magnetic fields up to 28 T at temperatures down to 0.35 K. The dHvA oscillations due to orbits from the Neck sections and "cubic box" of the Fermi surface were detected. We measured cyclotron effective masses of 0.5m$_0$ for Neck section of the FS. The dHvA frequencies as well as the cyclotron effective masses were calculated using the full potential Linear Muffin-Tin Orbital (LMTO) method within the Generalized Gradient Approximation. Comparison of the angular dependence of the dHvA frequencies with the band-structure calculations implies overall agreement with theoretical model, while one section could not be definitely identified.
PACS: 74.70.Ad, 74.60.Ec, 72.15.Gd, 78.20.Ci, 78.30.-j

Discovery of superconductivity in magnesium diboride [1] led to interest to superconductivity in other borides [2]. It was suggested by Matthias *et al.* [3] that the superconductivity in YB$_6$ and ZrB$_{12}$ was due to the effect of a cluster of light boron atoms. Much smaller isotope effect on $T_c$ for boron in comparison with Zr isotopic substitution suggests that the boron in ZrB$_{12}$ serves as inert background for the Zr driven superconductivity [4]. There has been little and controversial effort devoted to study basic superconducting and the electron transport properties of ZrB$_{12}$ (see Ref.5). In our recent study we demonstrated that the superfluid density of ZrB$_{12}$ displays unconventional temperature dependence with pronounced shoulder at T/T$_c$ equal to 0.65 [5]. We suggest that both two step $\lambda$(T) and linear H$_{c2}$(T) dependencies observed in ZrB$_{12}$ can be explained by two band BCS model with different not only superconducting gap but $T_c$ as well [5].

Recent electronic structure calculations of ZrB$_{12}$ [6,7] by self-consistent full-potential LMTO method have shown that the Fermi surface (FS) of ZrB$_{12}$ is composed of one open and one closed sheets. The knowledge of the experimental Fermi surface is critical for understanding superconductivity in this cluster compound. Until now there have been no direct experimental probes of the FS structure of ZrB$_{12}$. In this paper, we report the first study of the dHvA effect in ZrB$_{12}$ single crystals along with a detailed comparison with the predictions of band structure calculations. Four branches of dHvA frequencies are clearly resolved in our data and can be assigned to "Neck", "Dog's-bone" and "cuboidal box" orbits of the Fermi surface. The effective masses corresponding to the Neck orbits have been measured and compared with our band structure calculations. This showed a very large mass enhancement factor for this particular branch. As we expect the electron-phonon interaction to dominate the enhancement mechanisms, this leads to a large electron-phonon coupling for the "Neck" branch.

Under ambient conditions, dodecaboride ZrB$_{12}$ crystallizes in the *fcc* structure of the UB$_{12}$ type (space group *Fm3m*, *a*=0.74075 nm [5]). In this a rock salt type structure Zr is on Na and B$_{12}$ clusters on Cl sites. The boron atoms form B$_{12}$ cuboctahedral unit. All the dHvA atoms form B$_{12}$ cuboctahedral unit. All the dHvA measurements reported here have been performed on the same single crystal that had been previously studied in the electron transport, magnetic penetration depth and H$_{c2}$(T) [5]. The details of the sample preparation and characterization are presented elsewhere [5]. The sample dimensions are 0.5x0.5x2 mm$^3$, with its <110> axis parallel to its length. The critical temperature of the ZrB$_{12}$ samples, measured by RF susceptibility and $\rho$(T) was found to be T$_{c0}$=6.0 K, while the resistivity ratio, $\rho_{300K}/\rho_{6.5K}$=10, was rather low [5]. The dHvA oscillations (Fig.1a) were observed by measuring the torque with a capacitive cantilever technique [8]. The measurerements were performed on the M6 and M10 resistive magnets of the Grenoble High Magnetic Field Laboratory (of 23 and 28 Teslas respectively).

For these measurements, the sample was mounted in a holder equipped with a system allowing the sample to be rotated *in situ*. The angles were measured to a relative accuracy of at least one tenth of degree. In the measurements, <110> and <111> axis of the sample were parallel to the rotation axis, this allowed the magnetic fields to be rotated in the sample basal planes. The measurements were made at different temperatures in the range 0.35 K to 4.2 K with the sample immersed in a pumped $^3$He or $^4$He bath. The sample temperature was measured using a calibrated Cernox thermometer.

The oscillatory part of torque is given by $\widetilde{T} = -(1/F)\widetilde{M}_{//}(\partial F/\partial \theta)BV$ [8] where $\widetilde{M}_{//}$ - the oscillatory component of the parallel magnetization, $F$ is the dHvA frequency, $\theta$ is the orientation of the Fermi surface with respect to the applied filed $B$ and $V$ is the crystal volume. Here each dHvA frequency $F = S_F \hbar c/2\pi e$ is proportional to the extremal cross-section area $S_F$ of the Fermi surface. In a normal metal, $\widetilde{M}_{//}$ would be given by the usual Lifshitz-Kosevich formula. According to this formula, the dHvA amplitude $A$ of a fundamental frequency can be written as follows [8,9]:

$$A \propto B^{1/2} \left|\frac{\partial^2 S_F}{\partial k^2}\right|^{-1/2} \frac{\alpha m_c T/B}{\sinh(\alpha m_c T/B)} \exp(-\alpha m_c T_D/B),$$

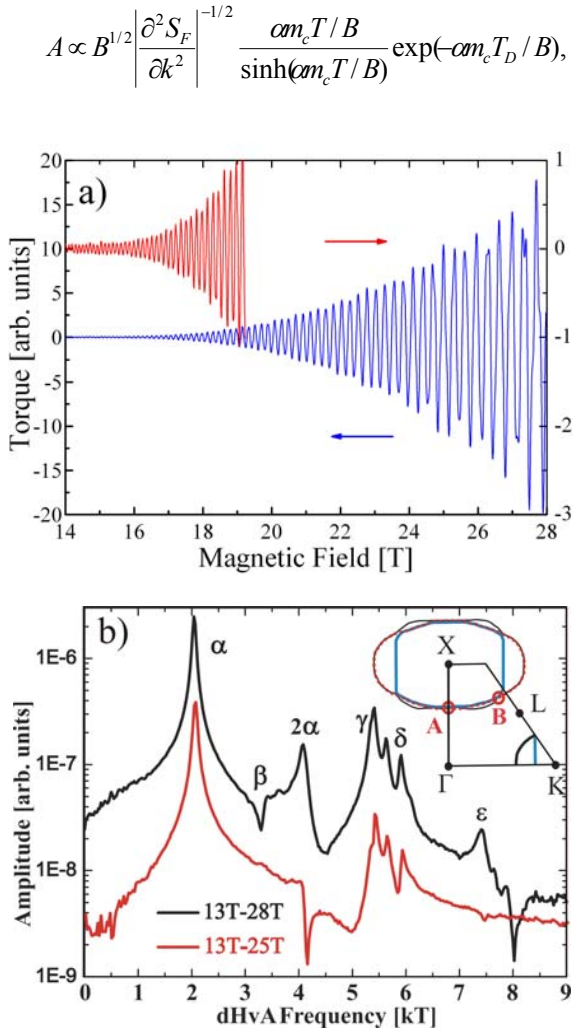

FIG.1. (Color online) Typical dHvA oscillations signal (a) and its Fourier spectrum (b) observed with magnetic field applied along <110> direction. Note second harmonic of $\alpha$ frequency. The peaks $\beta$ and $\varepsilon$ only appear for high maximum fields indicating its possible magnetic breakdown origin. Insert shows magnetic breakdown orbits through Dog's bone and cubic box sheet.

$$\alpha = \frac{2\pi^2 c k_B m_0}{e\hbar} = 14.693 (T/K) \quad (1)$$

here $m_c = \hbar^2(\partial S_F/\partial E)/2\pi$ is the cyclotron effective mass and $T_D = \hbar/2\pi k_B \tau$ is the Dingle temperature, inversely proportional to the quasiparticle lifetime $\tau$. We can try to estimate the residual $\tau$ from Drude formula for residual resistivity, $\rho(0) = 3/N_0 \tau v_F^2 e^2$, where we use measured $\rho(0)=1.8$ $\mu\Omega$cm [5] to obtain the Fermi surface averaged $\tau = 1.74 \cdot 10^{-14}$ sec. From this equation for $T_D$ we thus obtain very high FS averaged Dingle temperature of 70 K. Contrary to this estimation we observed rather large dHvA oscillations.

We show in Fig. 1 the typical dHvA data and the corresponding fast Fourier transform (FFT) spectrum for the magnetic field parallel to <110> for different maximum fields. As we can see, only weak harmonics of $2\alpha$ are observed for B//<110>, while for the other served for B//<110>, while for the other directions of B up to ten harmonics of $\alpha$ branch were seen. Fig. 2 shows the angular dependence of the dHvA frequencies (omitting the frequencies assigned as harmonics or its combinations). The circles represent the experimental data. The solid lines shows the results of the band-structure calculation described below. The FFT peaks denoted by $\alpha$, $\gamma$, $\delta$ and $\varepsilon$ are fundamental.

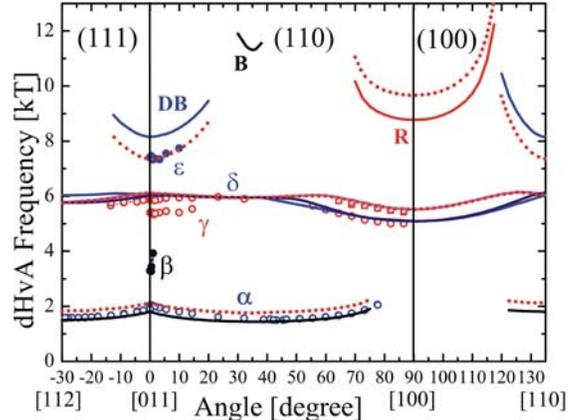

FIG. 2. (Color online) Experimental (symbols) and calculated (solid lines) dHvA frequencies of $ZrB_{12}$. DB, B and R denote the Dog's bone, Belly and Rosette orbits, respectively. Dotted lines shows result of calculations for $E_F$-0.16 eV.

The electronic structure of $ZrB_{12}$ was calculated using the full potential Linear Muffin-Tin Orbital (LMTO) method within the Generalized Gradient Approximation (GGA). The details of the calculation are presented elsewhere [7]. Our results are in very good agreement with the previously published band structure calculation [6]. The Fermi surface was computed over a 30×30×30 mesh in the first Brillouin zone (BZ). The corresponding extremal orbits from the first Cu-like hole FS sheet are shown in Fig.3a and those from the second "box"-like sheet in Fig.3b. The extremal areas, related to the dHvA frequencies through Eq.1, were obtained by slicing the calculated FS sheets perpendicular to the field direction, calculating the areas of all closed orbits and searching for the extrema among the slices. The cyclotron band mass for each extremal orbit is obtained by numerical differentiation $S_F$ versus $E$ with $\Delta E = 0.5$ Ry. For the calculation, the (100) plane was also considered for completeness.

The angular dependence of the dHvA frequencies (Fig.2) imposes strong constraints on the possible topologies of the FS and therefore allows the validity of the calculated FS (solid lines) to be verified. The assignment of the FFT peaks to the $ZrB_{12}$ FS sections was achieved by comparing the values of the frequencies obtained from FFT, and their angular dependencies. The solid lines in Fig.2, shows the excellent agreement between the *ab initio* electronic structure calculations and the measured one. The lowest frequency $\alpha$ branch between 1.5 and 2 kT corresponds to the Necks of the hole sheet 1. The branches $\delta$ in the vicinity of 6 kT are the signature of the nearly cubic boxes of sheet 2. The deviation from cube is evidenced by the splitting of the δ branch between the <111> and <100> field directions (Fig2). Also, the δ peak is split into two satellites δ and γ close to <110> direction. Apparently, this splitting is due to a small warping of the cubic box sheet not seen from calculations.

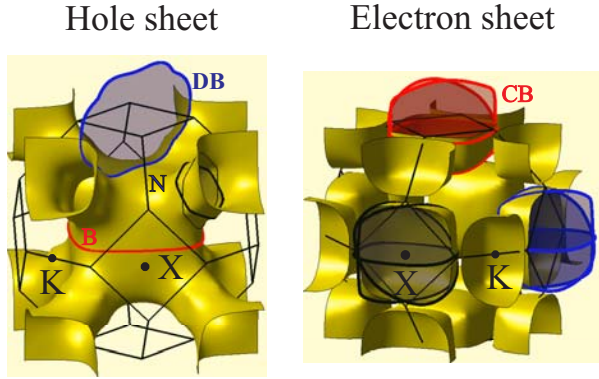

FIG.3. (Color online) Theoretical model of the Fermi surface of $ZrB_{12}$: electron and hole sheets considered in the dHvA branch calculation.

The highest measured frequency of 7.4 kT in the <110> direction, ε, does not seem to match any of the calculated orbits. The dHvA frequency of ε branch however is about 15% lower than results of calculated branch due to the "Dog's-bone" orbits. One possible explanation for this discrepancy is that the calculated orbit is too large (by approximately 0.8 kT), i.e. that the distance between the Necks is smaller in the real electronic structure than in the calculation. This could be due to a slight error in the calculation of the Fermi level.

In order to check this problem, we have done a calculation for a limited number of directions with $E_F$ shifted by -0.16 eV which was the estimated shift to bring the calculated "Dog's bone" branch over the experimental one in the [110] direction (dotted curves in Fig.2). However, the resulting shift of the Neck branch moves to another direction, while does not strongly affect the agreement with experimental points. Also, the size of the Necks which matches extremely well between the experiment and the calculation and that the size of the "Dog's-bone" orbit is related to the Neck size by the topology of the FS sheet as they both originate from the same band 1 sheet.

Another possible explanation for ε branch relies on the observation that two FS sheets are nearly or fully degenerate in several points of the BZ, i.e. the box-like FS sheets touch the Belly and the Necks. This is evidenced in inset in Fig.1b which shows the traces of the two FS sheets in the (110) plane: the Box and the Belly touch at the two high-symmetry points labeled "A" and the Box and the Necks nearly touch at the four "B" points (the calculated energy difference between the two bands at these points is less than 100 meV). With a frequency of 7.6 kT, the orbit shown in inset of Fig.1b by dotted line is the closer to the measured ε -branch. Also, the fact that closer to the measured ε -branch. Also, the fact that the observation the frequencies of the ε -branch depends on the applied magnetic field as shown in Fig.1b, moved us towards the magnetic breakdown scenario. There are different possibilities of reconnecting these magnetic breakdown orbits combining the two bands of the Fermi surface. Thus, observation of two additional peaks (due to different magnetic breakdown orbits) above ε peak, also support this magnetic breakdown conclusion.

While the β branch of dHvA frequencies just below 4 kT in the <110> direction may be an additional consequence of the magnetic breakdown of the "Dog's-bone" orbit, because magnetic field sensitivity of this peak, no frequencies below 4 kT could be obtained in reconnecting the orbits. The form of this peak is inverted relative to other ones, indicating a probable artifact.

The cyclotron effective masses $m_c$ were measured for the Neck orbit from the temperature dependence of the dHvA amplitude of α branch determined from FFT by performing field sweeps at different temperatures. According to Eq.1, the temperature dependence of the dHvA amplitude can be approximated as: $A/T$ = const/sinh($\alpha m_c T/B$). Figure 4 (top and right axis) shows the dependence of A/T versus T obtained for dHvA amplitude deduced from FFT. From the fitting of this dependence (solid curve), we determined the cyclotron effective mass equal to $0.5m_e$ for B//<111> where dHvA amplitude of α branch is strongest.

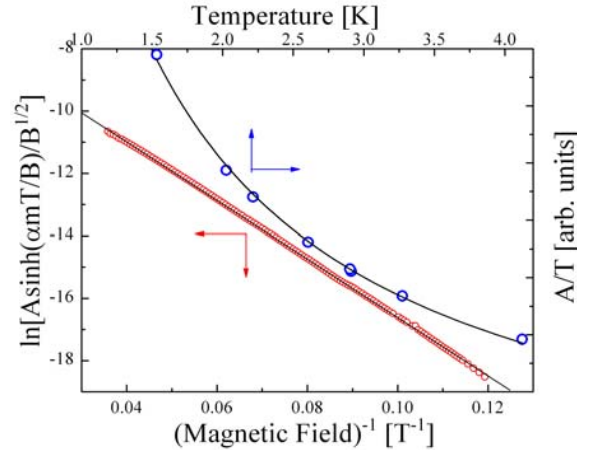

FIG.4. (Color online) Magnetic field dependence (left and bottom axis), for B//<111> of the Fourier amplitude of the dHvA α branch normalized by the field. Solid line is best fit using Eqs.1 of dHvA data with $T_D$ = 12.7K. Top and right axis: the temperature dependence of the dHvA amplitude of α oscillations at B//<111>. The fit (solid line) allows for the determination of the cyclotron mass $m_c=0.5m_e$.

The Dingle temperature can be obtained from the plot of $\ln[AB^{1/2}\sinh(\alpha m_c T/B)]$ versus $1/B$ according to Eq.1 (see left and bottom axis of Fig.4). For the B//<111>, the Dingle temperatures is 12.7 K for the α branch of dHvA oscillations. From the equation of $T_D$ we estimate the electron scattering time τ = 9.35 $10^{-14}$ sec almost five times larger than the one deduced from residual resistiv-ity. Unfortunately, the dHvA oscillations for β, γ, δ and ε sheets of the FS have too small intensities to allow an estimation of the $m_c$ and $T_D$ and comparison with a Neck data. Both cubic box and Neck sheets have similar very wide belt of the extremal orbits due to the form of those sheets. Nevertheless, the dHvA amplitude of α peak is almost ten times larger than the one for δ peak. One source of this difference could be a large difference in the scattering rate, i.e. $T_D$.

Our experimental values of $m_c$ for Neck orbit can be compared with the calculated band masses in Table I. The difference observed is the experimental mass enhancement due to the electron-phonon interaction [10]. We can calculate the electron-phonon coupling constant $\lambda_{ep}$, from $m_{exp}/m_{calc}=\lambda_{ep}+1$. The results show that the value of $\lambda_{ep}$ on Neck section is unusually large, 1.38. Optics also gave similar large electron-phonon constant (1.0) averaged for whole Fermi surface, while from specific heat very small $\lambda_{ep} = 0.2$ was observed [11]. Actually, similar large enhancement of electron-phonon interaction was observed for Neck sections of the FS of noble metals [10]. In the Neck region the hybridization of the wave functions and umklapp processes are very important on top of the k dependence of the matrix element of the electron – phonon interaction [10].

Fig.2 and Table I show that the values of the extremal areas $S_F$ found from dHvA effect and band structure calculations are in very good agreement. The differences between the measured $S_F$ values and calculated ones are not significant in many cases. The remaining discrepancy with band structure calculations concerns the absence of big Belly, Dog's bone and Rosette sections of sheet 1 dHvA orbits in the present study and the absence of extremal sections corresponding to *β* branch in calculated FS. The non-observation of these orbits could be due to the large $m_c$ for Belly sections and relatively short mean-free-path *l* on the hole sheet relative to electron sections. Thus future experiments with purer samples are essential for observation of Belly sections of the FS in a first BZ which may clarify the anisotropy of $\lambda_{ep}$ and validity of two-band model.

TABLE1. Summary of Fermi surface parameters for $ZrB_{12}$ sample along with theoretical prediction.

| Field | Orbit | $S_F^{exp}$,Å$^{-2}$ | $S_F^{calc}$,Å$^{-2}$ | $m_c^{calc}/m_0$ |
|---|---|---|---|---|
| <110> | Neck | 0.195 | 0.1722 | |
| | Box I | 0.49 | | |
| | Box II | 0.536 | 0.5765 | 0.47 |
| | DB | 0.712 | 0.7778 | 0.6 |
| <111> | Neck | 0.147 | 0.1375 | 0.21 |
| | Box I | 0.562 | 0.5677 | 0.44 |
| | Belly | | 1.0799 | 1.0 |
| <100> | Box I | 0.478 | 0.4852 | 0.4 |
| | Box II | 0.520 | 0.5272 | 0.4 |
| | Rosette | | 0.837 | 0.56 |

*In summary,* we have presented the first experimental study of the Fermi surface of the cluster superconductor $ZrB_{12}$, using dHvA effect. It is in excellent agreement with band structure calculations of the Fermi surface topology of this compound. The comparison of the experimental and calculated cyclotron mass shows unusually large electron-phonon interaction on Neck sections of the hole sheet of the Fermi surface. Results support observation of magnetic breakdown from Neck and Box sheets in fields above 25 T.

**Acknowledgments**

We would like to thank V.F. Gantmakher, R. Huguenin, D. van der Marel, I.R. Shein, for very useful discussions, V.B. Filipov, A.B. Liashchenko and Yu.B. Paderno, for preparation the $ZrB_{12}$ single crystal. This work was supported by the RAS Program: New Materials and Structures (Grant 4.13) and by the INTAS (Grant No. 2001-0617). A portion of this work was performed at the Grenoble High Magnetic Field Laboratory, which is supported by CNRS (Ref. SM0606). We acknowledge a support from Swiss National Science Foundation through grant 200020-109588 and the NCCR MaNEP for support of band structure calculation.